# Extreme Synergy in a Retinal Code: Spatiotemporal Correlations Enable Rapid Image Reconstruction

**Garrett T. Kenyon**

Physics Division, Los Alamos National Laboratory, Los Alamos, NM, USA 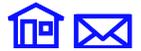

Over the brief time intervals available for processing retinal output, roughly 50 to 300 msec, the number of extra spikes generated by individual ganglion cells can be quite variable.  Here, computer-generated spike trains were used to investigate how signal/noise might be improved by utilizing spatiotemporal correlations among retinal neurons responding to large, contiguous stimuli.  Realistic correlations were produced by modulating the instantaneous firing probabilities of all stimulated neurons by a common oscillatory input whose amplitude and temporal structure were consistent with experimentally measured field potentials and correlograms.  Whereas previous studies have typically measured synergy between pairs of ganglion cells examined one at a time, or alternatively have employed optimized linear filters to decode activity across larger populations, the present study investigated a distributed, non-linear encoding strategy by using Principal Components Analysis (PCA) to reconstruct simple visual stimuli from up to one million oscillatory pairwise correlations extracted on single trials from massively-parallel spike trains as short as 25 msec in duration.  By integrating signals across retinal neighborhoods commensurate in size to classical antagonistic surrounds, the first principal component of the pairwise correlation matrix yielded dramatic improvements in signal/noise without sacrificing fine spatial detail.  These results demonstrate how local intensity information can distributed across hundreds of neurons linked by a common, stimulus-dependent oscillatory modulation, a strategy that might have evolved to minimize the number of spikes required to support rapid image reconstruction.

Keywords: synchrony, synchronization, gamma oscillation, spatiotemporal correlation, computer model, retina, retinal ganglion cell, neural code, rate-code, temporal-code, multi-unit activity, linear filter

## Introduction

The conditions under which retinal ganglion cells transmit visual signals synergistically remains a topic of considerable debate (Latham & Nirenberg, 2005; Schneidman, Bialek, & Berry, 2003; Schnitzer & Meister, 2003).  Measurements of synergy between small groups of neurons have found evidence for everything from redundancy (Gawne & Richmond, 1993; Puchalla, Schneidman, Harris, & Berry, 2005; Warland, Reinagel, & Meister, 1997) to statistical independence (Nirenberg, Carcieri, Jacobs, & Latham, 2001; Panzeri, Schultz, Treves, & Rolls, 1999; Reich, Mechler, & Victor, 2001; E. T. Rolls, Franco, Aggelopoulos, & Reece, 2003) to both modest (Dan, Alonso, Usrey, & Reid, 1998) as well as more substantial levels of synergy (deCharms & Merzenich, 1996; Gat & Tishby, 1999 ; Hirabayashi & Miyashita, 2005; Riehle, Grun, Diesmann, & Aertsen, 1997; Samonds, Allison, Brown, & Bonds, 2004; Singer, 1999; Vaadia et al., 1995).  Several studies have looked specifically for synergy among larger ensembles (Bezzi, Diamond, & Treves, 2002; Frechette et al., 2005; Narayanan, Kimchi, & Laubach, 2005; Stanley, Li, & Dan, 1999), yet it remains an open question as to how a non-linear code based on spatiotemporal correlations between hundreds of ganglion cells, corresponding to hundreds-of-thousands of coactive neuron pairs, might convey local pixel-by-pixel intensity information that could not be obtained by analyzing the same spike trains individually, especially over the short time scales—approximately 50 to 300 msec—available for interpreting retinal signals (Bacon-Mace, Mace, Fabre-Thorpe, & Thorpe, 2005; Kirchner & Thorpe, 2006; Edmund T. Rolls, Tovee, & Panzeri, 1999; Thorpe, Fize, & Marlot, 1996).

Here, Principal Components Analysis (PCA) was used to explore how local pixel intensities could be encoded in a non-linear manner by exploiting oscillatory spatiotemporal correlations among co-activated ganglion cell pairs.  Coherent oscillations in response to diffuse stimuli have been reported in many species, including frog (Ishikane, Gangi, Honda, & Tachibana, 2005; Ishikane, Kawana, & Tachibana, 1999), mudpuppy (Wachtmeister & Dowling, 1978), rabbit (Ariel, Daw, & Rader, 1983; Yokoyama, Kaneko, & Nakai, 1964), cat (Doty & Kimura, 1963; Laufer & Verzeano, 1967; Neuenschwander, Castelo-Branco, & Singer, 1999; Neuenschwander & Singer, 1996; Steinberg, 1966), monkey (Frishman et al., 2000; Laufer & Verzeano, 1967) and humans (De Carli et al., 2001; Wachtmeister, 1998).  Short segments of computer-generated spike train data, from 25 to 400 msec in duration, were used to simulate a retinal patch containing up to one thousand output neurons, represented by a single ganglion cell type whose receptive field centers precisely tiled a square rectilinear grid. Both background and stimulated firing rates were chosen so as to encompass much of the measured dynamic range of cat retinal ganglion cells in response to diffuse, sinusoidally varying stimuli (Troy & Enroth-Cugell, 1993). Realistic spatiotemporal correlations were generated by simultaneously modulating the instantaneous firing probabilities of all stimulated cells using a common, oscillatory waveform whose amplitude and temporal structure was consistent with both optic tract recordings (Doty & Kimura, 1963; Laufer & Verzeano, 1967; Steinberg, 1966) and with single and multi-unit correlation functions (Ishikane et al., 1999; Neuenschwander & Singer, 1996). Over a nearly 16-fold range of stimulated firing rates, the present results show that without sacrificing fine spatial detail, oscillatory pairwise correlations can support rapid pixel-by-pixel reconstructions of large, contiguous visual stimuli that are far superior to analogous reconstructions based on Poisson-distributed event trains that contained the same average number of spikes at each stimulus intensity,.

The present findings imply that oscillatory correlations consistent with those reported between retinal ganglion cells in several species can substantially augment the stimulus information conveyed by a simple rate-code, the latter being defined by a code in which local pixel intensity is conveyed entirely by the number of spikes produced by the corresponding ganglion cell. Neither purely spatial correlations in the number of spikes, nor purely temporal correlations within the individual spikes trains, could by themselves account for the superior quality of the image reconstructions obtained here by fully exploiting both spatial and temporal correlations simultaneously. Spatiotemporal correlations mediated performance levels on an ON/OFF pixel discrimination task that, if instead mediated by an independent rate-code, would have required approximately four times as many spikes to achieve similar accuracy. By distributing local intensity information across an extended neighborhood of contiguously activated cells, the present findings indicate that rapid image reconstructions can be accomplished using far fewer impulses than would otherwise be required.

## Methods

### Computer-Generated Spike Trains

*Overview*: The procedures used here for generating artificial spike trains are similar to those previously described (Kenyon, Theiler, George, Travis, & Marshak, 2004; Stephens, Neuenschwander, George, Singer, & Kenyon, 2006). Starting with a power spectrum containing a single Gaussian peak and characterized by three parameters, denoting the amplitude, width, and central frequency, respectively, the signal was transformed back to the time domain by randomly choosing the phases of the individual Fourier components. Parallel sets of artificial spike trains with realistic spatial-temporal correlations were then constructed by assuming an array of binary event generators, using the above time series to simultaneously modulate their instantaneous firing probabilities. Although on any given trial, every cell activated by the stimulus had the same time-dependent firing probability, the spike-trains themselves nonetheless varied considerably from cell-to-cell due to the underlying stochastic process. A mathematically similar procedure has been used to model local field potentials based on a collection of random phase oscillators (Diaz, Razeto-Barry, Letelier, Caprio, & Bacigalupo, 2007), although here the order of operations has been reversed; in the present study spike trains were constructed from an oscillatory Peri-Stimulus-Time-Histogram (PSTH) that contained multiple frequencies added in random phase, as opposed to starting with oscillatory spike trains with different frequencies and random phases that were subsequently combined into a multi-unit PSTH.

*Mathematical Details*: An oscillatory time series of duration, $T$, and temporal resolution, $\Delta t$, was constructed by first defining the discrete frequencies, $f_k$:

$$f_k = \frac{k}{T}, 0 \leq k < \frac{T}{\Delta t} \tag{1}$$

in terms of which the discrete Fourier coefficients were defined as follows:

$$C_k = \exp(2\pi i r_k)\exp\left(\frac{(f_k - f_0)^2}{2\sigma^2}\right) \tag{2}$$

where $f_0$ is the central oscillation frequency, here set to 80 Hz, $\sigma$ is the width of the spectral peak in the associated power spectrum, here set to 10 Hz, and $r_k$ is a uniform random deviate between 0 and 1 that randomized the phases of the individual Fourier components (generated by the Matlab® intrinsic function RAND). A different random sequence was generated each stimulus trial. The coefficients, $C_k$, were used to convert back to the time domain using the discrete inverse Fourier transform (generated by the Matlab® intrinsic function IFFT), so that the firing rate, $R_n$, at each time step, $t_n = n \Delta t$, was given by:

$$R_n = I_i \frac{1}{N}\sum_{k=1}^{N-1} C_k e^{-2\pi i f_k t_n} + R_0 \tag{3}$$

where the $R_n$ denotes only the real part of the sum on the RHS of eq. 3 and $N = T/\Delta t$. The scale factor, $I_i$, and the constant offset, $R_0$, were set so that the mean stimulated firing rate, $<R^{(i)}>$, and the standard deviation, $\sigma_R^{(i)}$, were given by the following relations:

$$\left\langle R^{(i)} \right\rangle = R_0\left(1 + \frac{(2^i - 1)}{4}\right) \tag{4}$$

$$\sigma_R^{(i)} = R_0 \frac{(2^i - 1)}{8} \tag{5}$$

for $i = \{0,1,2,3,4,5\}$ specifying different stimulus intensities. For $i = 0$, we have simply $I_0 = 0$, $R_n = R_0 = 25$ ips, for all $n$. For $i > 0$, however, the situation was complicated by the fact that negative values of $R_n$ were truncated to zero, making it necessary to determine the constants $I_i$ and $R_0$ empirically via an iterative procedure. This was accomplished by recalculating $<R^{(i)}>$ and $\sigma_R^{(i)}$ after setting all values of $R_n<0$ to zero and adjusting $I_i$ and $R_0$ so that equations 4 and 5 were satisfied. Negative values of $R_n$ were then again truncated at zero and the process repeated until the discrepancy from the exact equality expressed by equations 4 and 5 was less than 0.5%.

Once determined, the time series defined by $R_n$ was used to generate oscillatory spike trains via a pseudo-random process:

$$S_n = \theta(R_n \Delta t - r) \qquad (6)$$

where $R_n \Delta t$ is the probability of a spike in the $n^{th}$ time bin, $\theta$ is a step function, $\theta(x<0) = 0$, $\theta(x\geq 0) = 1$, and $r$ is again a uniform random deviate. In the limit that $R_n \Delta t << 1$, the above procedure reduces to a rate-modulated Poisson process. The same time series, $R_n$, was used to modulate the firing probabilities of each element contributing to the artificially generated multi-unit spike train, thus producing oscillatory spatiotemporal correlations due to common input.

The range of oscillatory modulations employed here produced periodic temporal correlations between co-activated cells that were consistent with electrophysiological recordings across a variety of vertebrate retinas. The multi-unit activity (MUA) obtained by combining spikes across all stimulated cells exhibited periodic deflections that increased sharply in response to stimulus intensity (Figure 1, column **MUA**) in a manner that was qualitatively consistent with optic nerve recordings in both cats and primates following full-field stimulation (Doty & Kimura, 1963; Laufer & Verzeano, 1967; Steinberg, 1966). Likewise, cross-correlation functions extracted from the computer-generated spike trains (Figure 1, column **Cross-Correlation**), obtained by averaging over all stimulated cell pairs, were again qualitatively similar, in terms of relative peak amplitude, shape and width, to corresponding experimental measures from cats (Neuenschwander et al., 1999; Neuenschwander & Singer, 1996) and frogs (Ishikane et al., 2005; Ishikane et al., 1999) in response to large or looming stimuli, respectively. Considerable evidence suggests that at least some types of retinal neurons can oscillate coherently in response to diffuse stimuli. The spatiotemporal correlations employed here are within the scope of these reports.

### Estimating Pairwise Correlations

*Method 1: SYNC.* Single-trial cross-correlations were first estimated by counting the number of synchronous events between each pair of neurons, defined as the occurrence of a spike in both cells in the same 1 msec time bin. The expected number of synchronous events due to chance, computed separately on each trial, was subtracted from the actual count, yielding a correlation estimate that was positive, on average, for pairs of cells that tended to fire together and distributed about zero for cells that fired randomly with respect to each other. Mathematically, the SYNC-based cross-correlation, $X_{ij}$, was given by

$$X_{ij} = \sum_{t_k} \left[ S_i(t_k) - \bar{S}_i \right] \left[ S_j(t_k) - \bar{S}_j \right] \qquad (7)$$

where $S_i$ denotes the spike train of the $i^{th}$ neuron, either 0 or 1 at the corresponding time step, $t_k$, and $\bar{S}_i$ is the single-trial average.

Autocorrelations were computed identically, such that each event was treated as "synchronous" with itself, thereby preserving rate-coded information along the diagonals and setting the maximum correlation amplitude.

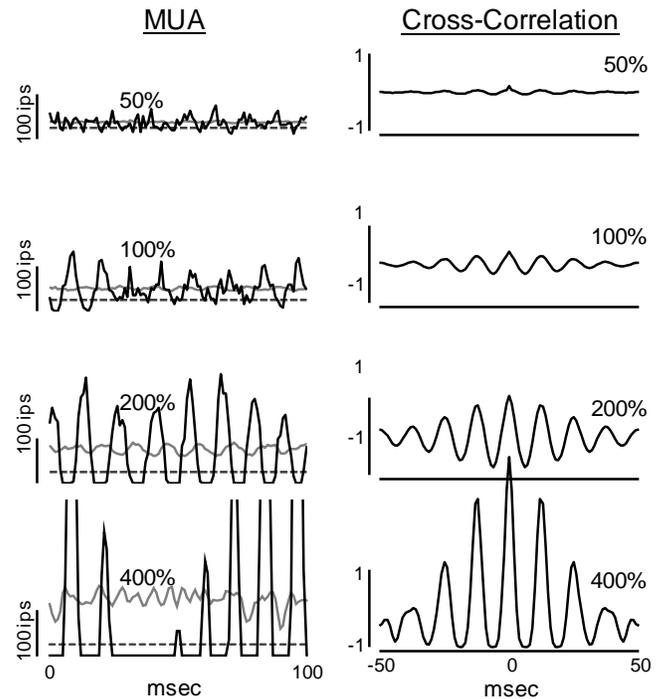

Figure 1. Temporal structure of computer-generated spike trains. MUA: Instantaneous firing rate (1 msec bins) combining spike trains from all 16×16 stimulated cells (black lines) exhibit oscillatory modulations that increased with stimulus intensity, indicated on each panel. When averaged over 100 independent trials, oscillatory structure in the instantaneous firing rate largely disappeared (gray lines), due to the lack of time-locking to stimulus onset. Baseline firing rate indicated by dashed lines. Cross-Correlation: Average pairwise cross-correlation between foreground pixels, expressed as a fraction of baseline. Both the increase in the mean firing rate, as well as the amplitude and shape of the coherent oscillatory modulations, fell within the range of published values.

*Method 2: γMUA\*.* Pairwise correlations were estimated using a second method that depended on whether relative spike timing was consistent with the oscillatory component the local multi-unit activity (γMUA\*). An estimate of γMUA\* was obtained for each 1 msec time step by first summing all spikes in the neighborhood of the target cell, out to a maximum radius of 4 pixels. Each spike from a given remote cell within the integration neighborhood was weighted by the inverse of the distance to the target cell, with distance measured as the radius of the square circumference centered on the target cell and passing through the remote cell. This measure of distance ensured that the contribution from each square perimeter was independent of the number of cells at that radius, preventing the estimate of the local MUA from being dominated by the outermost rings. The local MUA was band pass filtered using a rectangular window between 60 and 100 Hz, exclusive of the endpoints, yielding the oscillatory component of the local MUA, denoted here as γMUA*, where the asterisk serves as a reminder that this quantity is merely analogous to the band-pass filtered multi-unit activity that would be measured experimentally. The γMUA*, in turn, was used to weight each event in the remote and target spike trains, giving positive weight to those events occurring at the peaks of the γMUA* and negative weights to events occurring in the troughs. By design, the weighted sum of random events would equal zero, on average, because the γMUA* itself has zero mean.

An estimate of the oscillatory cross-correlation was given by the sum over the product of all pairs of spikes in the two trains, with each event weighted by the γMUA* at the target cell. Mathematically, the γMUA*-based cross-correlation, $\Gamma_{ij}$, was given by

$$\Gamma_{ij} = \sum_{t_k, t_l} \gamma_i(t_k) S_i(t_k) \gamma_j(t_l) S_j(t_l) \qquad (8)$$

where $\gamma_i$ is a short-hand for γMUA* at the $i^{th}$ location and the sum is over all pairs of events, or equivalently, over all pairs of time steps, $t_k$ and $t_l$, regardless of their relative timing.

Autocorrelations were computed similarly, so that the diagonal terms in the pairwise correlation matrix reflected the square of number of events produced by each cell, modulated by the degree of overlap with the γMUA*. For Poisson distributed activity, diagonal terms were distributed about zero, on average, except for the contribution of each cell to its own γMUA*, which effectively ensured some residual autocorrelation.

One disadvantage of the above procedure was that strongly anti-correlated spike trains could be assigned large, positive cross-correlation strengths, but this was not an issue in the current study as our computer-generated spikes trains were never explicitly anti-correlated. The main advantage of the above procedure was is that it allowed correlations strengths to be estimated from only a few events based on their relative timing, regardless of whether such events were synchronous or not. Attempts to estimate oscillatory correlation strengths using conventional Fourier analysis (Kenyon, Harvey, Stephens, & Theiler, 2004) were unsuccessful, due to the extremely limited number of events available from very short spike train segments. By comparison, the γMUA* permitted the instantaneous phase of the stimulus-dependent oscillation to be reliably estimated by averaging over a small neighborhood, yielding a meaningful estimate of the correlation strength from as little as a single pair of spikes.

**Stimulus Reconstruction**

*Rate-Based Reconstructions*: Stimuli were first reconstructed by setting the magnitude of each pixel proportional to the logarithm of the number of spikes generated by the corresponding cell, normalized by the expected number of spikes due to baseline activity. The maximum reconstructed pixel intensity was determined by the maximum number of spikes fired by a cell in response to the largest stimulus intensity across all trials. The number of trials was 100 unless otherwise noted. The minimum number of spikes was truncated at the average baseline level, so that the minimum pixel intensity was equal to zero. Since our computer-generated spikes trains had no antagonistic surrounds, rate deviations below baseline were due entirely to noise. From the distributions of the number of spikes produced by stimulated vs. unstimulated cells, an ideal threshold was determined for each stimulus intensity, allowing optimal classification of pixels as either ON or OFF (see *Ideal Observer Classification* below). Sub-threshold pixels values were set to zero, which had the effect of suppressing the background, making the reconstructed images easier to interpret.

*SYNC- and γMUA*-Based Reconstructions*: Stimuli were also reconstructed using the logarithm of the normalized first principal component, or largest eigenvector, of the pairwise correlation matrix computed using either the SYNC or γMUA* methods described above. For all stimulus intensities, the normalization factor was equal to average pixel value of the first principal component computed in response to baseline activity. The first principal component was computed using the Matlab® intrinsic function EIGS, with the pairwise correlation matrix, either $X_{ij}$ or $\Gamma_{ij}$, replaced with the explicitly symmetrical construction $\vec{X}^T\vec{X}$ or $\vec{\Gamma}^T\vec{\Gamma}$. To resolve an overall sign ambiguity, the average value of all stimulated pixels was required to be positive—all pixels values were multiplied by −1 if this condition was violated—except when the intensity was zero. The maximum pixel value was determined by the largest value obtained in response to the largest intensity across all stimulus trials. Background activity was again suppressed by using the distributions of stimulated and unstimulated pixel values to compute an ideal threshold for each stimulus intensity and setting all subthreshold pixels to zero.

As a control, stimuli were also reconstructed using only the diagonal terms of the γMUA*-based correlation matrix, preserving information encoded in the temporal alignment of individual spike trains with the γMUA* but discarding terms that depended on the relative timing between spike trains. As a final control, stimuli were reconstructed after replacing the γMUA*-based weighting of each spike by a uniform weight of one, thereby preserving spatial correla-tions in the total number of spikes while discarding information regarding relative spike timing or oscillatory temporal structure.

### Ideal Threshold Determination

Regardless of the reconstruction method employed, an ideal discrimination threshold could be determined by examining the distributions of pixel values representing either 1) stimulated vs. unstimulated activity (i.e. ON vs. OFF) or 2) two different non-zero stimulus intensities. Theoretically, the maximum percentage of pixels that could be correctly discriminated is inversely related to the degree of overlap between the two distributions (Duda, Hart, & Stork, 2001). If the distributions overlapped completely, the maximal theoretical performance on the pixel discrimination task would be no better than chance (50% correct). On the other hand, if the distributions were entirely non-overlapping, the maximum theoretical performance on the discrimination task would be perfect (100% correct). Between these two extremes, corresponding to distributions that partially overlap, maximum theoretical performance on the pixel discrimination task, $P$, expressed as a fraction of trials correctly classified, is given by the following formula:

$$P = (2 - A_{overlap})/2 \qquad (9)$$

where $A_{overlap}$ denotes the total area of the overlap between the two distributions and the maximum value of $A_{overlap}$ is normalized to one. Error bars on the estimated values of $P$ were determined by assuming the pixel values to either side of the Bayes discriminator obeyed binomial statistics. Error bars were always negligible and were omitted on semi-logarithmic plots.

## Results

Based on trial-to-trial variability, an independent rate-code can be quite noisy, even under circumstances where the evoked activity is a factor of two or more above unstimulated baseline levels (Barlow & Levick, 1969). On a pixel-by-pixel basis, the intrinsically low signal/noise conveyed by an independent rate code can be illustrated by examining idealized responses to a series of simple visual stimuli presented at various intensities (Figure 2, column IMAGE). Computer generated spike trains, 100 msec in duration and constructed using a binary process operating close to the Poisson limit (step size equal to 1 msec), were used to model both background activity, here specified by a mean value of 25 impulses per second (ips), as well as stimulated activity, which ranged from 25% to 400% above baseline levels. These firing parameters encompassed both measured baseline rates (Troy & Robson, 1992) as well as much of the dynamic range of Y ganglion cells in the cat retina in response to diffuse drifting gratings presented at up to 100% luminance contrast and temporally modulated at 4 Hz (Troy & Enroth-Cugell, 1993). Although retinal neurons can fire in a highly reproducible manner in response to temporally complex or high-contrast stimuli (Passaglia & Troy, 2004; Reich, Victor, Knight, Ozaki, & Kaplan, 1997; Reinagel & Reid, 2000; Uzzell & Chichilnisky, 2004), responses to moderate intensity, slowly varying stimuli often exhibit considerable trial-to-trial variability (Barlow & Levick, 1969).

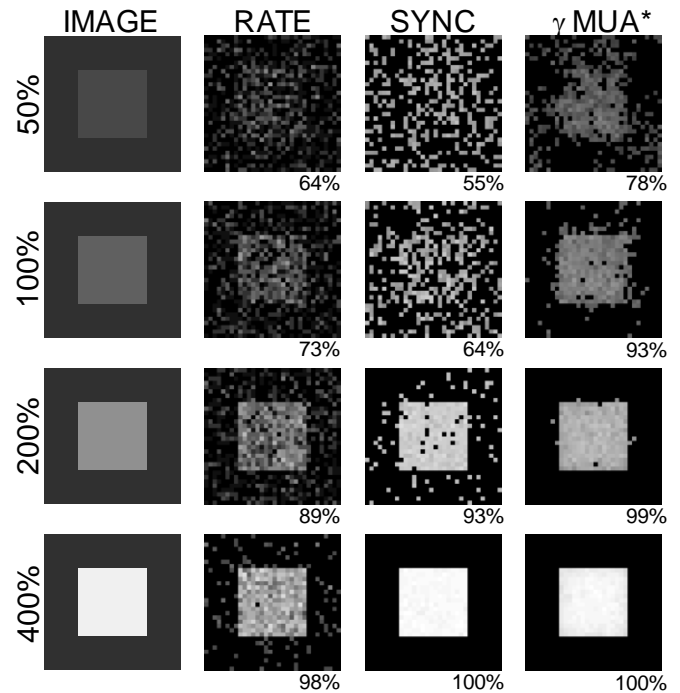

Figure 2. Single-trial reconstructions. Computer generated pikes trains 100 msec in duration, with a baseline rate of 25 ips, were used to simulate firing activity in a 32×32 retinal patch. IMAGE: Stimuli were 16×16 uniform square spots. Intensity, indicated to the left of each panel, denoted the increase in firing rate relative to baseline and, if present, the RMS amplitude of the oscillatory modulation. RATE: Representative reconstructions based on the logarithm of the number of spikes. SYNC: Reconstructions based on the logarithm of the largest principle component of the (32×32)×(32×32) correlation matrix computed from the number of synchronous events between each pair of spike trains relative to chance. γMUA*: Reconstructions computed as for SYNC but with correlations estimated by weighting events in each pair of spike trains by the oscillatory component of the local multi-unit activity at each target cell (γMUA*) and summing over all weighted event pairs. For all three reconstruction methods, the ideal threshold for classifying individual pixels as either ON or OFF was determined for each intensity level and subthreshold pixels were set to zero. Ideal theoretical performance indicated to bottom right of each panel. Correlations dramatically improved stimulus reconstructions at all intensities.

Single-trial, rate-based reconstructions were obtained by setting the intensity of each pixel to the logarithm of the number of spikes generated by the corresponding neuron, normalized relative to the average number of spikes due to baseline activity (Figure 2, column RATE). A logarithmic scale was used to maintain sensitivity over the full dynamic range. Background activity was suppressed by choosing a discrimination threshold that maximized the percentage of pixels correctly classified as either ON or OFF (Duda et al., 2001) and setting all sub-threshold pixel values to zero. If the computed discrimination threshold was less than zero, a value of zero was used instead. The percentage of correct ON/OFF classifications, averaged across all trials, is shown at the bottom right of each reconstruction. Representative examples, here and in subsequent panels, corresponded to individual trials on which the percentage of correct classifications was closest to the multi-trial average. Doubling the firing rate relative to baseline, from 25 ips to 50 ips over a period of 100 msec, supported performance levels of only 73% for the ON/OFF classification task, compared to an expected performance level of 50% for purely random assignments. Such poor performance can be readily explained, since doubling the firing rate produced on average only 2.5 extra spikes per pixel, corresponding to a signal/noise of only 1.6, consistent with the trial-to-trial variability exhibited by cat ganglion cells over similar intervals (Barlow & Levick, 1969).

Synergistic encoding schemes, in which an oscillatory modulation was used to simultaneously vary the firing rates of all activated neurons, were investigated as well. The applied oscillatory modulation produced periodic spatiotemporal correlations between all neuron pairs underneath the stimulus, information that could be used to identify groups of cells whose outputs could be pooled to achieve superior signal/noise. To mimic the contrast-dependence of the coherent oscillations measured experimentally, which increase in strength with luminance contrast (Neuenschwander et al., 1999), the RMS amplitude of the oscillatory modulation, expressed as a fraction of the average response to the stimulus, was set equal to the fraction by which the mean stimulated firing rate exceeded the baseline rate.

Correlation strengths were evaluated in one of two ways. First, the number of simultaneous events between each pair of cells was summed over the 100 msec analysis interval and expressed as the difference from the expected number of synchronous events due to chance. Second, a measure that was more sensitive to periodic structure and which took account of temporal correlations at all relative delays (i.e. not just synchronous spikes falling in the same 1 msec time bin) was used. Briefly, an estimate of the local multi-unit activity (MUA) at each target cell was obtained by summing all spikes out to a given radius, typically 4 pixels, such that the influence of each event decreased as one over the distance to the target cell. The MUA was bandpass filtered using a rectangular window from 60 Hz and 100 Hz, yielding an estimate of the oscillatory component of the local multi-unit activity, denoted $\gamma MUA^*$. The $\gamma MUA^*$ was then used to weight each spike, emphasizing those events occurring during peaks of the local oscillation and discounting those events falling in the troughs. The sum over the product of all weighted spike pairs provided an estimate of the oscillatory correlations between any two cells that was much more sensitive than conventional Fourier analysis when only a few events were available.

Depending on how pairwise correlations were estimated, either using synchrony (SYNC) or the oscillatory component of the local multi-unit activity ($\gamma MUA^*$), stimuli were reconstructed by computing the first principal component, or largest eigenvector, of the corresponding pairwise correlation matrix. Using synchrony to measure pairwise correlation strengths (Figure 2, column SYNC), the first principal component yielded a noticeable improvement over the reconstructions mediated by an independent rate-code, but only at higher stimulus intensities. If the $\gamma MUA^*$ was instead used to estimate pairwise correlation strengths (Figure 2, column $\gamma MUA^*$), the largest singular vector yielded dramatic improvements in stimulus reconstruction over an independent rate code across a range of intensities spanning nearly 4 $\log_2$ units (16-fold). This improvement was quantified by measuring relative performance on the ON/OFF pixel classification task. Whereas a doubling of the firing rate supported only 73% correct classifications using an independent rate-code, an oscillatory correlation-code supported an average of 92% correct classifications for the same average increase in firing activity.

Spatiotemporal correlations supported not only improved performance on the ON/OFF pixel discrimination task but also mediated improved discrimination between different stimulus intensities, separated by factors of 2. As a function of stimulus strength, the distributions of foreground pixel values were much better separated when periodic spatiotemporal correlations were used to reconstruct the input image as opposed to the rate-based reconstructions (Figure 3, top row). The black dotted line indicates the distribution of foreground pixel values in the absence of stimulation. Progressively lighter shades corresponded to increasing stimulus strengths, except for the highest intensity, which was denoted by a solid black line. The x-axis was scaled logarithmically to better separate the distributions at low stimulus intensities (normalized units). Vertical dashed lines indicate the discrimination thresholds used in the ON/OFF classification task (note some thresholds were zero and thus off scale). All distributions were normalized to unity (the apparent area was distorted by the logarithmic x-scale). Distributions of foreground pixel values obtained from the rate-based reconstructions overlapped considerably, especially between lower stimulus strengths, implying poor intensity discrimination. At larger stimulus strengths, the distributions obtained from the SYNC-based reconstructions were reasonably well separated, but only the distributions derived from $\gamma MUA^*$-based reconstructions were relatively well separated at nearly all intensities.

For each pair of reconstructed images, obtained in response to different stimulus strengths, a discrimination threshold was chosen that allowed optimal classification of each foreground pixel with respect to the two intensities being compared. Performance on the intensity discrimination task, measured as the percentage of correctly classified pixels assuming either intensity was equally likely, was plot-ted as a function of the relative intensity difference between the two stimuli, measured in $\log_2$ units (Figure 3, middle row). For the rate-based reconstructions, performance improved monotonically both as a function of the relative intensity difference and as the reference intensity—or weaker of the two intensities—was increased (Figure 3A, RATE, reference intensity was smallest for the bottom-right-most curve, largest for upper-left-most curve). The improved performance with increasing reference intensity reflected the fact that higher firing rates supported larger signal/noise. For the correlation-based reconstructions, performance on the intensity discrimination task was substantially better for most combinations of relative and reference stimulus strengths (Figure 3, columns SYNC and γMUA*). This was especially true for the more sensitive γMUA*-based correlation measure, for which performance levels exceeded 90% for intensity differences between 1 to 3 $\log_2$ units, whereas an intensity difference of 3 or more $\log_2$ units was required to ensure the same level of performance using the rate-based reconstructions.

Similar results were obtained when the data were reanalyzed to extract the discrimination threshold, or minimum intensity difference, required to support a given level of performance on the intensity discrimination task (Figure 3, bottom row). Compared to the rate-based reconstructions, discrimination thresholds were uniformly lower for the correlation-based reconstructions, particularly when pairwise correlation strengths were estimated using the γMUA* method. At the 90% performance level, discrimination thresholds for the γMUA*-based reconstructions ranged from 1 to 2.5 $\log_2$ units lower than for the rate-based reconstructions. Thus, on a pixel-by-pixel basis, measures of spatiotemporal correlations, particularly those sensitive to periodic structure, permitted greatly superior discrimination between different stimulus intensity levels.

The dramatic improvements in signal/noise mediated by spatiotemporal correlations presumably reflect the aggregate contributions from tens-of-thousands of stimulated cell pairs. The number of elements in the non-symmetric, γMUA*-based pairwise correlation matrix was $N^2 = 1,048,576$, representing 1,024 potential sources of information per pixel. However, only ¼ of the cells were activated by the stimulus, so a lower estimate of the number of useful pairwise correlations would be $1/16^{th}$ this value, representing 256 additional sources of information per pixel. Assuming that the sheer number of pairwise correlations was responsible for the observed improvement in signal/noise, it follows that smaller patches, containing smaller numbers of cell pairs, should support proportionately smaller improvements in signal/noise compared to an independent rate code.

To test this prediction, reconstructions were performed using progressively smaller retinal patches, down to a minimum size of 4×4 neurons stimulated by a 2×2 spot, for which the number of potentially useful pairwise correlations, including autocorrelations, was only 4 per pixel. The percentage of correct classifications on the ON/OFF pixel discrimination task was plotted as a function of the total size of the retinal patch for the 4 largest stimulus intensities employed (Figure 4). For the rate-based reconstructions (solid line), the percentage of correct classifications was insensitive to the size of the retinal patch, as expected for independent encoders. For the γMUA*-based reconstructions (dashed line), performance increased monotonically with patch size, saturating at approximately 24×24 total neurons, or at 12×12 stimulated cells. This latter number is in reasonable agreement with estimates of the size of redundant cell neighborhoods in the salamander retina (Puchalla et al., 2005).

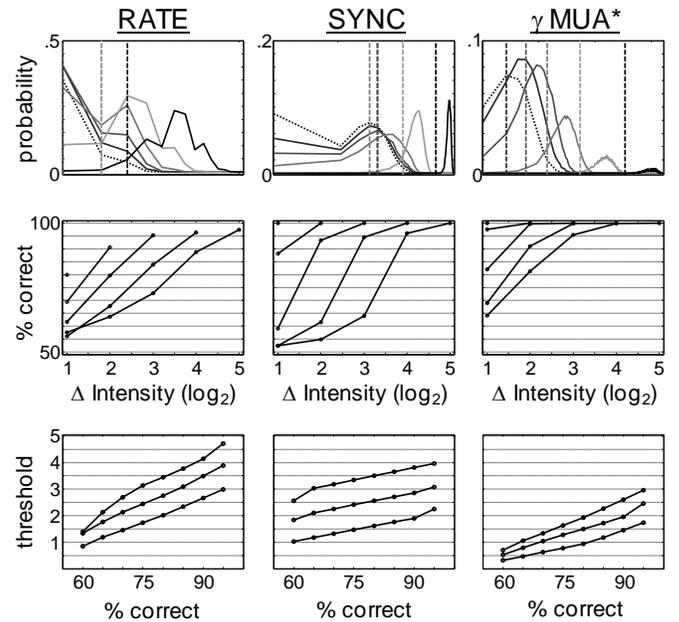

Figure 3. Discrimination between different stimulus intensities. Reconstruction method indicated at the top of each column. *Top Row*: Distribution of foreground pixel values (semilogarithmic scale). Dotted line denotes the baseline distribution. Lighter shades correspond to increasing stimulus intensity (from 25% to 200%). The maximum intensity (400%) denoted by a solid black line. Vertical dashed lines indicate discrimination thresholds used in the ON/OFF pixel class task, with shading matched to corresponding intensity. Distributions from γMUA*-based reconstructions are more widely separated than distributions for SYNC- or RATE-based reconstructions. *Middle Row*: Percentage of correct pixel discriminations as a function of intensity difference, plotted in $\log_2$ units. The reference intensity, which increased from the bottom right-most to the top left-most curve on each plot, was matched by an upward shift in the percent correct, reflecting improved signal/noise at higher firing rates. Performance was markedly superior for the γMUA*-based reconstructions. *Bottom Row*: Intensity discrimination threshold in $\log_2$ units, as a function of the percentage of correct classifications. Discrimination thresholds were 1.5 to 3 $\log_2$ units lower for the γMUA*-based reconstructions (curves shown for the three lowest reference intensities, which increased from top to bottom).

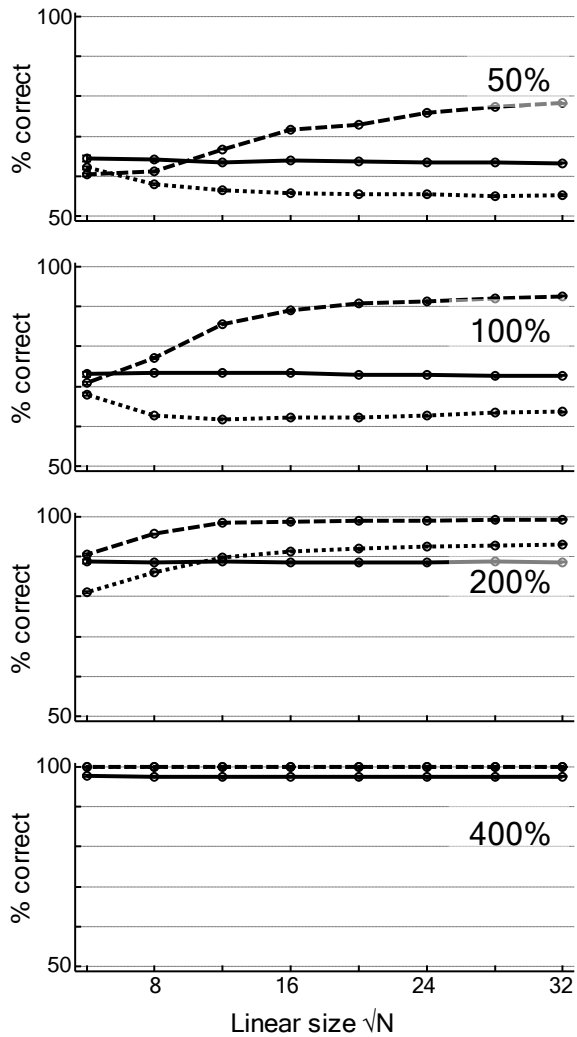

Figure 4. Dependence on number of pairwise correlations. Performance on the ON/OFF pixel classification task was plotted as a function retinal patch diameter, with the correlation matrix constructed from all cell pairs. Stimulus intensity indicated on each panel. The precision of the rate-based reconstructions (solid lines) was independent of patch size. At small to moderate stimulus intensities, performance mediated by the $\gamma$MUA*-based reconstructions (dashed lines) rose steeply as a function of the total number of cells, saturating for patch sizes approximately 24×24. At the highest intensity, superior $\gamma$MUA*-based reconstructions required only a few cells. The quality of the SYNC-based reconstructions (dotted lines) exhibited a more complex dependence on patch size, due to noise in the underlying estimates of pairwise correlation strength, which was the dominant factor at low stimulus intensities. Number of trials: {100, 100, 200, 300, 400, 600, 800, 1000} for patch diameters {32, 28, 24, 20, 16, 12, 8, 4}, respectively.

Paradoxically, for SYNC-based reconstructions at lower stimulus intensities, increasing patch size led to lower performance on the ON/OFF pixel classification task, presumably because the first principal component was dominated by noise in the off-diagonal terms of the pairwise correlation matrix, swamping rate-coded information distributed along the diagonal. On the other hand, even for very small patches, the performance mediated by the $\gamma$MUA*-based reconstructions was either equivalent to, or for higher stimulus intensities measurably better than, that mediated by the rate-based reconstructions. Overall, these results support the hypothesis that for weak to moderate pairwise correlation strengths, the improvements in signal/noise provided by the $\gamma$MUA*-based reconstructions required simultaneous processing of tens- to hundreds-of-thousands of cell pairs.

As a consequence of trial-to-trial variability, individual spike trains provided unreliable information about local pixel intensity. It was possible to substantially reduce trial-to-trial variability, however, by pooling over groups of correlated cells. Remarkably, such pooling did not obscure fine spatial detail, as each term in the pairwise correlation matrix, which was based on a sum over the product of event pairs, was always proportional to the local firing activity. The preservation of fine spatial detail was demonstrated implicitly in the ability to reproduce sharp edges in the correlation-based reconstructions of uniform square spots (Figure 2). When individual pixels were randomly deleted from the stimulus, thus creating an arbitrary complex shape, $\gamma$MUA*-based reconstructions continued to preserve spatial structure at the level of individual pixels (Figure 5).

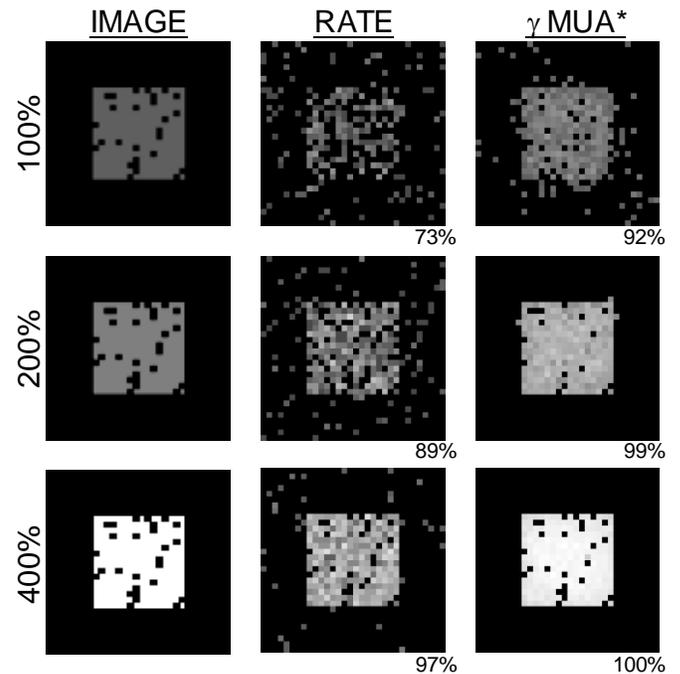

Figure 5. Preservation of fine spatial detail. Explanation of panels as in Figure 2 with column SYNC omitted. Stimuli were again 16×16 square spots but with 10% of the pixels randomly deleted. The percentage of correct ON/OFF pixel classifications was essentially identical as for uniform square spots. The largest principal component mediated greatly improved signal/noise without sacrificing fine spatial detail. Similar improvements in signal/noise via conventional spatial averaging would have required precise foreknowledge of which pixels had been deleted.

The first principal component integrated information distributed across many cells while maintaining a high fidelity representation of the pixel-by-pixel intensity. To obtain comparable improvements in signal/noise via conventional spatial averaging—without sacrificing fine spatial detail—would have required a separate template for every possible combination of deleted pixels. On the other hand, γMUA*-based correlation analysis required only the information available on single trials from the spike trains themselves. The present results demonstrate that even over very short time scales, from tens to hundreds of milliseconds, enough information may be available in the pairwise correlation matrix, especially when referenced to the local multi-unit activity (i.e. γMUA*), to construct such filters on the fly, thereby conferring the main advantages of spatial averaging (i.e. improved signal/noise) without sacrificing fine spatial detail.

Both psychophysical and electrophysiological data suggest that 50 to 300 msec are required to process visual scenes (Bacon-Mace et al., 2005; Kirchner & Thorpe, 2006; Edmund T. Rolls et al., 1999; Thorpe et al., 1996). The lowest estimates reflect the latency of image-specific responses recorded from individual neurons in IT compared to latencies recorded in V1 (Edmund T. Rolls et al., 1999). The longest estimates are based on the upper range of mean saccade reaction times for an image classification task in which only correct trials are considered (Kirchner & Thorpe, 2006) and falls within the range of inter-saccade intervals measured in humans and other primates (Martinez-Conde, Macknik, & Hubel, 2004). Consistent with the observed rapidity of visual processing, oscillatory spatiotemporal correlations were found to support improved stimulus reconstructions over time scales as short as 25 msec (Figure 6). Superior reconstructions were obtained from 25 msec spike trains even though the individual frequency components used in estimating the γMUA* at each target cell were quantized at 40 Hz intervals, so that periodic structure could at best be only poorly resolved. Correlation-based reconstructions were particularly accurate at higher intensities, with the first principal component becoming nearly perfect as the stimulus strength approached maximum values.

Retinal ganglion cells typically fire most strongly during the transient phase of the response immediately following stimulus onset, settling to a lower plateau level of activity after several tens of msec (Cleland, Levick, & Sanderson, 1973; Creutzfeldt, Sakmann, Scheich, & Korn, 1970). While the largest intensities modeled here are consistent with the highest levels of activity measured in cat Y ganglion cells in response to diffuse, sinusoidally modulated gratings (Troy & Enroth-Cugell, 1993), abrupt light steps produce transient peaks of activity that equal or exceed the maximum response amplitudes considered here (Cleland et al., 1973; Creutzfeldt et al., 1970). When assessing the quality of rate- vs. correlation-based reconstructions extracted from very short time spike train segments—on the order of 25 msec—the results derived here from higher stimulus intensities may therefore be more relevant.

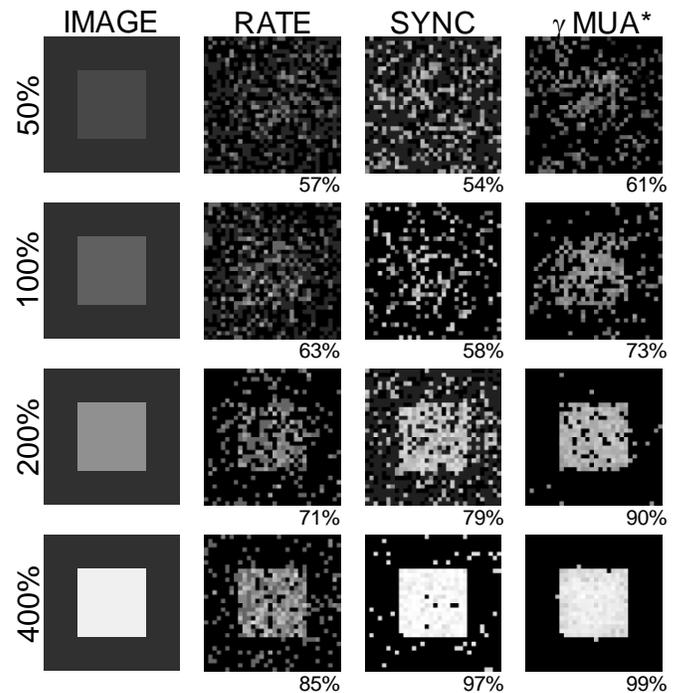

Figure 6. Reconstructions from 25 msec spike trains. Explanation of panels as in Figure 2. The γMUA*-based reconstructions were again greatly superior to those mediated by an independent rate code, despite the necessarily low resolution of the underlying frequency components. SYNC-based reconstructions were also superior at very high stimulus intensities.

To characterize how the quality of the rate-, SYNC- and γMUA*-based reconstructions depended on the length of the analysis window, optimal theoretical performance on the ON/OFF pixel classification task was plotted as a function of spike train duration, ranging from 25 to 400 msec (Figure 7). As expected, performance for all three reconstruction methods declined as the length of the analysis window was decreased, yet even for the shortest spike trains tested, the γMUA*-based reconstructions mediated substantially superior ON/OFF pixel discrimination across a nearly 16-fold range of intensities. At a stimulus intensity of 100%, representing a doubling in the firing rate and an RMS oscillatory amplitude that was two times baseline, γMUA*-based reconstructions supported performance levels exceeding 90% in less than 100 msec, whereas nearly 400 msec were required to achieve the same level of performance using rate-based reconstruction methods. Thus, γMUA*-based reconstructions required approximately 1/4$^{th}$ the number of events to achieve accuracy comparable to an independent rate-code. Given that the signal/noise in response to slowly varying stimuli presented at low to moderate contrast generally improves as the square root of the number of events (Barlow & Levick, 1969), the present findings imply that utilizing spatiotemporal correlations approximately doubles the potential signal/noise, an im-provement comparable to quadrupling the number of spikes. At the very weakest intensity tested, however, performance on the ON/OFF pixel discrimination task failed to reach 80% correct for any of the reconstruction methods used here, implying that additional mechanisms may be contributing to the perception of low contrast features.

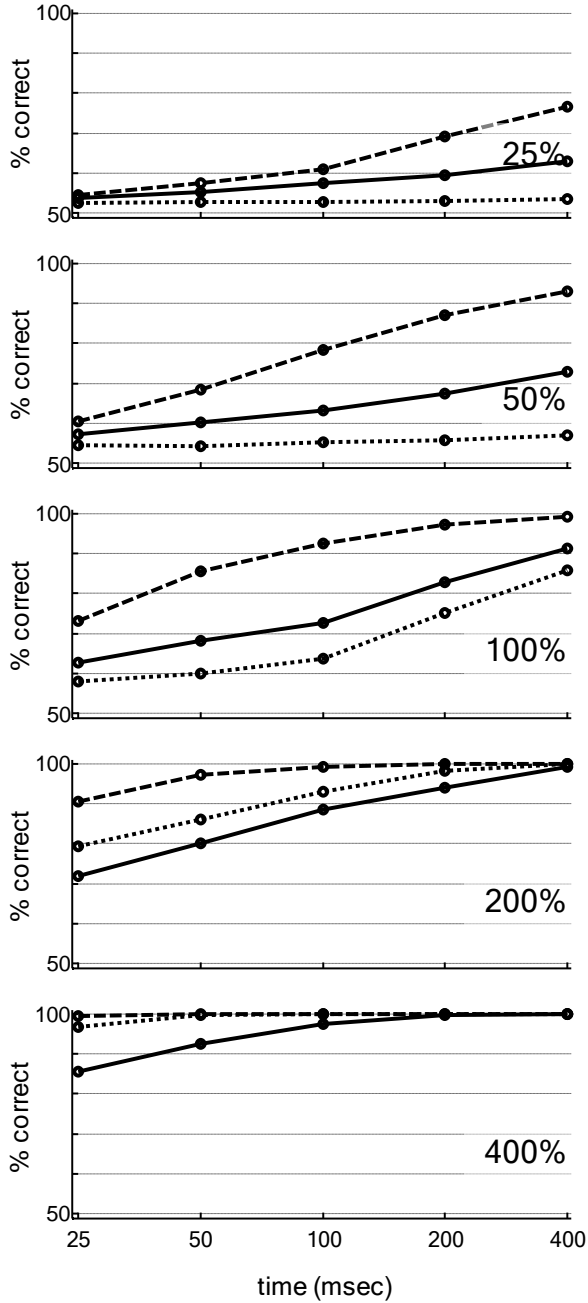

Figure 7. Reconstructions vs. spike train duration. Reconstruction quality assessed as the percentage of pixels correctly classified as either ON or OFF using an optimal discrimination threshold. Stimulus intensity indicated on each panel. Note logarithmic time scale. Rate-based reconstructions (solid lines) took substantially longer than $\gamma$MUA*- (dashed lines) and, at high intensities, SYNC-based reconstructions (dotted lines), to achieve comparable accuracy.

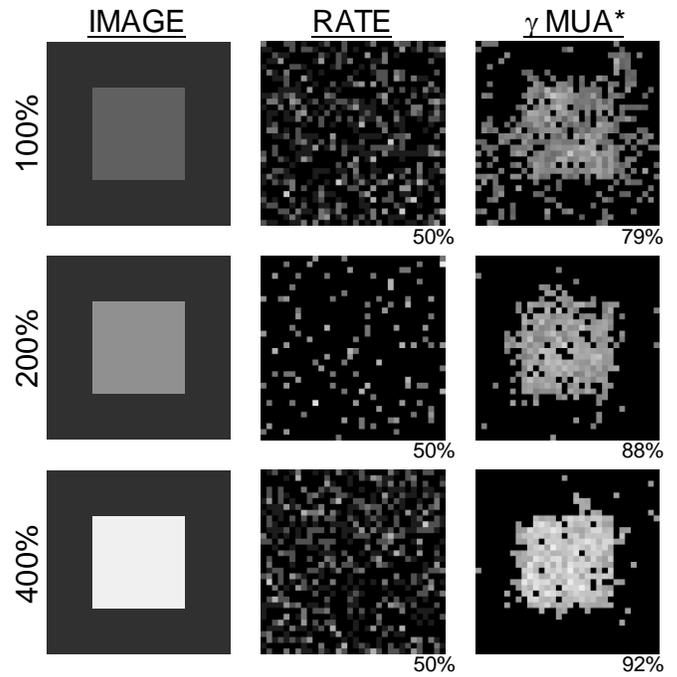

Figure 8. Reconstructions in the absence of rate-coded information. Explanation of panels as in Figure 5. Firing rates were held fixed at baseline levels. Stimulus intensities indicated the RMS amplitude of the oscillatory modulation relative to baseline. Rate-based reconstructions were indistinguishable from chance. Oscillatory spatiotemporal correlations conveyed less information in the absence of proportionate increases in the firing rate but could nonetheless support levels of accuracy in the range of 80% to 90% at higher stimulus intensities.

For a given intensity, coherent oscillations increase with stimulus size (Neuenschwander et al., 1999), suggesting that the spatiotemporal correlations evoked by large or diffuse stimuli might support accurate reconstructions even when the firing rate is little changed from baseline. To test this hypothesis, the quality of the $\gamma$MUA*-based reconstructions for various RMS oscillatory amplitudes was evaluated while keeping the time-averaged firing rate fixed at the baseline level (Figure 8). Although signal/noise was degraded in the absence of rate-coded signals, strong, coherent oscillations alone could nonetheless support reasonably accurate reconstructions, allowing the intensity of a large, contiguous region to be discriminated even at locations where the firing rate remained at or near baseline levels.

To parse the relative factors underlying the ability of oscillatory spatiotemporal correlations to mediate superior reconstructions, a series of experiments were performed to isolate the contributions from various sources of information, such as purely temporal modulations, spatial correlations, changes in the mean firing rate, and the common reference signal provided by the $\gamma$MUA* (Figure 9). Reconstruction quality was assessed as optimum performance on the ON/OFF pixel classification task, plotted as a function of stimulus intensity.

By themselves, strong oscillatory modulations reduced the trial-to-trial variability, regardless of whether the precise temporal structure was ignored. To isolate the contribution from reduced variability, a comparison was made between 1) rate-based reconstructions that used temporally-modulated spike trains, generated via the same mathematical process used to produce $\gamma$-band oscillations, (Figure 9, top panel, gray-solid line) and 2) rate-based reconstructions using non-modulated spike trains (black-solid line). Oscillatory temporal structure, by reducing the variability in the number of spikes generated over the 100 msec trial, yielded only a small improvement in signal/noise. Although retinal spike trains can exhibit significant temporal structure (Reich et al., 1997; Rodieck, 1967; Troy, Schweitzer-Tong, & Enroth-Cugell, 1995), these findings indicate that temporal structure alone cannot account for the superior quality of the $\gamma$MUA*-based reconstructions demonstrated here (dashed black line).

Conversely, stimulus-dependent changes in the mean firing rate contributed substantially to the quality of $\gamma$MUA*-based reconstructions, the latter being severely degraded in the absence of proportionate changes in the average number of spikes evoked at different stimulus intensities (Figure 9, top panel, dashed gray line). Although $\gamma$MUA*-based reconstructions were adversely affected by the loss of rate-coded information, oscillatory spatiotemporal correlations—even with the firing rate held at baseline—supported performance levels roughly equivalent to those mediated by a non-temporally-modulated rate-code (gray vs. solid black lines). Thus, oscillatory spatiotemporal correlations, if distributed across a sufficiently large number of cell pairs, can convey useful pixel-by-pixel intensity information even if the average firing rate is held constant.

A related question relates to whether purely spatial correlations, independent of the contribution from precise or oscillatory temporal structure, could have contributed meaningfully to the superior quality of the $\gamma$MUA*-based reconstructions. To address this question, a new correlation matrix was computed which ignored precise temporal structure while preserving spatial correlations. Specifically, pairwise correlations were again estimated by summing over all event pairs, taking one event in turn from each train regardless of relative lag, but no longer weighting each spike by the instantaneous value of $\gamma$MUA* at the target cell. Spike trains beneath the stimulus were still subjected to oscillatory temporal modulations identical to those used to support superior $\gamma$MUA*-based reconstructions. However, the recomputed correlation matrix discarded all dependence on the precise temporal structure, keeping only the spatial correlations in the total number of spikes. The resulting reconstructions, based on purely spatial correlations, were nearly identical to the rate-based reconstructions obtained from non-temporally modulated spike trains (Figure 9, middle panel, solid gray vs. solid black lines). Thus, spatial correlations in the number of events, even between cells subjected to a common oscillatory modulation, did not contribute meaningfully to the superior qual-ity of the γMUA*-based reconstructions. This finding is reasonable given that spatial correlations in the total number of events was not explicitly tied to the stimulus intensity, as were the amplitudes of the common oscillatory modulations applied to all stimulated cells. Indeed, reconstructions that utilized only spatial correlations were identical in quality to rate-based reconstructions using temporally-structured spike trains (Figure 9, solid gray lines, middle vs. top panels), indicating that trial-to-trial fluctuations in the total number of events generated by separate spike trains conveyed no useful information regarding stimulus intensity.

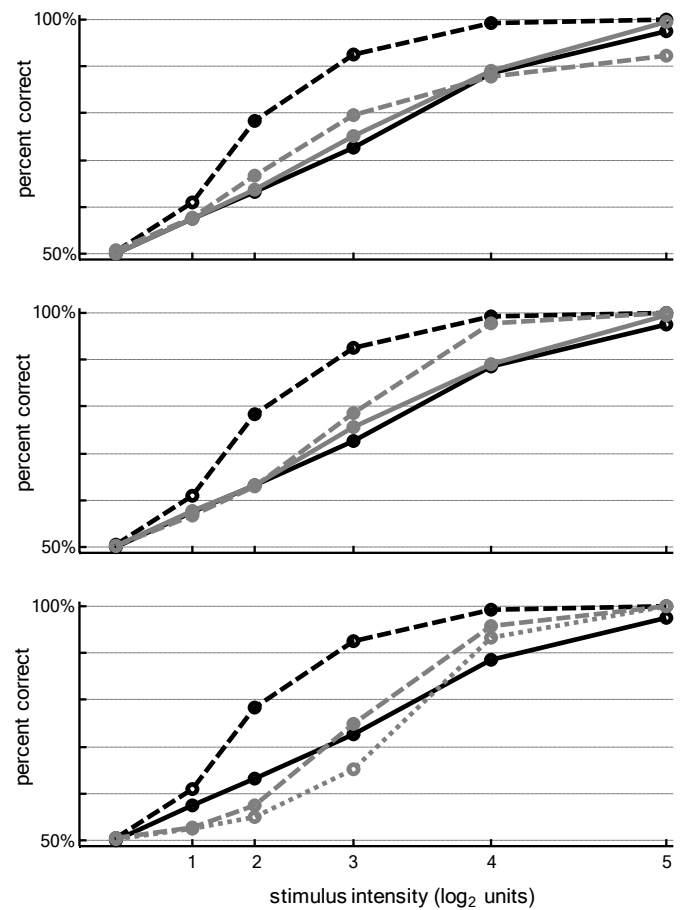

Figure 9. Relative contributions of spatial and temporal factors. Reconstruction quality plotted as optimal ON/OFF pixel classification performance vs. stimulus intensity. Solid and dashed black lines in all three panels give performance of the previously described rate- and $\gamma$MUA*-based reconstructions, respectively. Top Panel: Rate-based reconstructions using spike trains subjected to $\gamma$-band temporal modulations (solid gray line) were only slightly improved relative to a non-temporally-modulated event trains, despite reduced trial-to-trial variability. Alternatively, $\gamma$MUA*-based reconstructions were noticeably degraded in the absence of proportionate firing rate increases (dashed gray line), although spatiotemporal correlations alone could support performance similar to that mediated by a pure rate-code. Middle Panel: Incorporating spatial correlations into rate-based reconstructions that used temporally-modulated spike trains yielded no improvement (solid grey line), as spatial fluctuations in spike counts were not stimulus related. Ignoring off-diagonal terms in the pairwise correlation matrix degraded $\gamma$MUA*-based reconstructions (dashed gray line), highlighting the importance of spatial-correlations. Bottom Panel: When pairwise correlations were assessed using standard Fourier analysis, without reference to the local multi-unit activity, the corresponding reconstructions were degraded (dashed gray line), underlining the importance of the $\gamma$MUA* in providing a common reference signal. The performance mediated by SYNC-based reconstructions is show for comparison (dotted gray line).

Our findings further suggest that precise temporal structure alone, even when referenced to the local phase of the multi-unit activity, fails to account for the superior quality of the γMUA*-based reconstructions. Modified γMUA*-based reconstructions were obtained using only the diagonal terms in the pairwise correlation matrix, thus preserving the autocorrelation of each spike train with the local phase of the multi-unit activity but discarding all cross-correlations. Such diagonal-only reconstructions were markedly inferior to the standard γMUA*-based reconstructions that utilized off-diagonal correlations as well (Figure 9, middle panel, dashed gray vs. dashed black lines), except at high stimulus intensities where differences may have been masked by saturation effects. Thus, it can be inferred from the above results that spatial correlations make a critical contribution to the superior quality of γMUA*-based reconstructions, but only when both spatial correlations and precise temporal structure are simultaneously taken into account.

Finally, experiments were conducted to estimate the contribution to the superior quality of the γMUA*-based reconstructions arising from the local multi-unit activity, which provided a reference signal for estimating pairwise correlations. A new pairwise correlation matrix was computed by performing a Fourier analysis of each spike train segment. Correlation strengths were estimated by the peak in the cross-power spectra between 60Hz and 100Hz (given by product of the Fourier amplitudes from the two spike trains) multiplied by the cosine of the difference in relative phase at the peak frequency. The dependence on the relative phase ensured that synchronous oscillatory activity corresponded to a positive correlation, whereas activity that was out of phase corresponded to a negative pairwise correlation. Taking the average of all such products over the interval 60Hz to 100Hz, roughly corresponding to the area under the peak, yielded equivalent results (data not shown). Reconstructions based on a pairwise cross-power spectra analysis, which ignored alignment with local multi-unit activity, were greatly inferior to the γMUA*-based reconstructions (Figure 9, bottom panel, dashed gray vs. dashed black curves) but were slightly better than SYNC-based reconstructions (dotted gray curve). The poor reconstructions derived from a Fourier-based analysis of pairwise cross-power spectra reflects the difficulty of estimating oscillatory correlations from a very small number of events. The γMUA*, by averaging over a local neighborhood containing multiple cells, yielded an estimate of the instantaneous phase of the common oscillation that allowed a meaningful weight to be assigned to every inter-spike interval. Thus, by taken account of the known temporal structure linking cells responding to the same contiguous region, the oscillatory component of the local multi-unit activity permitted an estimate of the correlation strength to be obtained for each pair of events.

## Discussion

As independent encoders, retinal ganglion cells appear poorly designed to convey precise pixel-by-pixel intensities on short time scales. In response to sustained or smoothly varying stimulation, especially at low to moderate contrast, retinal spike trains can be quite noisy, exhibiting considerable trial-to-trial variability (Barlow & Levick, 1969). The problem of extracting information from retinal spike trains is especially critical given that subjects require only 50-300 msec to process visual scenes (Bacon-Mace et al., 2005; Hung, Kreiman, Poggio, & DiCarlo, 2005; Kirchner & Thorpe, 2006; Edmund T. Rolls et al., 1999; Thorpe et al., 1996), a time interval over which the standard deviation in the number of spikes generated by a typical retinal ganglion cell is often a substantial fraction of the mean (Barlow & Levick, 1969). How then, in the face of such pronounced trial-to-trial variability, can the perceived acuity of visual experience be explained?

The findings presented here suggest an explanation based on extreme synergy, such that the messages conveyed by individual retinal ganglion cells can only be properly interpreted within the context of the concurrent firing activity across extended neighborhoods roughly commensurate in size to the classical antagonistic surround. Using computer-generated spike trains modulated by realistic coherent oscillations, the present findings demonstrate that spatiotemporal correlations can support pixel-by-pixel estimates of the input stimulus intensity that in many cases are dramatically superior to those mediated by uncorrelated, rate-matched spike trains. The proposed synergistic encoding mechanism is intrinsically non-linear and thus represents an alternative to reconstruction strategies based on optimal linear filters (Stanley et al., 1999; Warland et al., 1997), in which the contribution from each spike is incorporated independently of any other spikes.

Here, using single-trial estimates of the pairwise correlation matrix, the contribution from each spike to any given reconstruction was explicitly weighted by its time of occurrence relative to other firing events; draw either from the same or from surrounding cells. Spikes whose relative timing was aligned with the oscillatory component on the local multi-unit activity were more likely to be evoked at peaks of the common temporal-modulation and thus more likely to convey reliable intensity information. Non-aligned spikes, on the other hand, were more likely to reflect background activity or noise, and thus exerted proportionately smaller influence on the final reconstruction. As an illustration, imagine that each spike train consisted of two entirely separate event types; randomly generated background spikes, whose rate was independent of stimulus intensity, and precisely correlated spikes that accurately encoded the local stimulus intensity. An effective decoding strategy would be to first label every pair of spikes as either "correlated" or "uncorrelated", and then to estimate local stimulus intensities using only "correlated" pairs of spikes. Although the present reconstruction algorithm, based on PCA, did not work as cleanly as in this idealized example, the underlying principle is nonetheless very similar.

The hypothesis that coherent oscillations represent detailed stimulus properties was motivated in part by extrapolations of marginal entropy measurements to large ensembles of weakly correlated ganglion cells, which indicate that the observed patterns of activity may correspond to distinct attractors (Schneidman, Berry, Segev, & Bialek, 2006). Here, the use of PCA permitted the information distributed across large neural ensembles to be recombined at each pixel, rapidly accomplishing significant improvements in signal/noise without sacrificing fine spatial detail.

Previous attempts to measure synergy between pairs of neurons in the visual pathway have obtained contradictory results (Dan et al., 1998; Gawne & Richmond, 1993; Hirabayashi & Miyashita, 2005; Nirenberg et al., 2001; Puchalla et al., 2005; Reich et al., 2001; E. T. Rolls et al., 2003; Samonds et al., 2004; Singer, 1999). No previous experimental study, however, has directly considered the precise pixel-by-pixel information encoded across tens-of-thousands of non-linear, pairwise correlations processed simultaneously. Consistent with measurements of synergy between pairs of ganglion cells responding to natural images (Nirenberg et al., 2001; Puchalla et al., 2005), we found relatively little synergy across small retinal patches containing only a few ganglion cells, unless the pairwise correlations were both very strong and explicitly stimulus-dependent. Rather, it was only when patch size increased sufficiently so that large numbers of moderate or weak stimulus-dependent correlations could be processed simultaneously that significant amounts of additional information became available. Other studies have shown that pairwise correlations among large numbers of retinal ganglion cells do not improve estimates of global stimulus parameters based on population-coded averages (Frechette et al., 2005). The present study, however, relates only to the precise pixel-by-pixel information distributed across simultaneously recorded neurons. When estimating the global stimulus properties of a large object, such as speed, the improvements in signal/noise obtained by averaging over many cells are so substantial that spatiotemporal correlations are unlikely to yield detectable improvements and indeed may limit the advantages of spatial averaging (Shadlen & Newsome, 1994), although the stimulus information encoded by the correlations themselves can compensate for this effect (Kenyon, Theiler et al., 2004). However, to average over cells in a manner that preserves fine spatial detail requires prior knowledge of the precise dimensions of the stimulus, whereas the first principal component of the pairwise correlation matrix could be directly extracted on single trials from the spike trains themselves without any prior knowledge of the input image.

It is uncertain how time-dependent firing rates, such as transient-sustained periods of activity evoked by step onsets (Cleland et al., 1973; Creutzfeldt et al., 1970) or temporally structured responses to high contrast drifting gratings (Reich et al., 1997), might have influenced the fidelity of the rate-based reconstructions. It is also uncertain whether substantially improved rate-based reconstructions might have been possible using the responses to dynamic image sequences, which are known to evoke highly reproducible firing patterns (Passaglia & Troy, 2004; Reinagel & Reid, 2000; Uzzell & Chichilnisky, 2004). Improvements in signal/noise resulting from temporal structure alone, such as has been reported to underlie improved estimates of object (Frechette et al., 2005) and whole-image (Nemenman, Lewen, Bialek, & Stevenninck, 2006) motion parameters, was not comprehensively addressed. Nonetheless, it was shown that the temporal structure imposed by strong oscillatory modulations, ignoring the concomitant spatial correlations, was insufficient to support high fidelity rate-based reconstructions, despite the presence of small improvements that resulted from reduced trial-to-trial variability, or larger improvements that followed from direct analysis of the precise temporal structure referenced to the $\gamma MUA^*$.

It has also been proposed that rapid, pixel-by-pixel intensity information is transmitted to the brain via spike latency codes (Van Rullen & Thorpe, 2001). In principle, latency codes could be multiplexed with long-rage spatiotemporal correlations, as coherent oscillations do not impose any particular order of firing within any given cycle. Similarly, coherent oscillations are not inconsistent with evidence that synchronous spikes can encode higher spatial resolution (Schnitzer & Meister, 2003), as the short-range synchrony postulated to underlie spatial hyperacuity may involve different, although possibly overlapping, synaptic mechanisms (Meister & Berry, 1999) to those responsible for long-range synchrony (Kenyon et al., 2003). Finally, optimal linear filters, especially those which take account of overlapping ganglion cell surrounds (Stanley et al., 1999; Warland et al., 1997), would likely provide additional intensity information to the explicitly non-linear, synergistic encoding scheme examined here.

Coherent oscillations between retinal neurons have been reported in many vertebrate species (Ariel et al., 1983; De Carli et al., 2001; Doty & Kimura, 1963; Ishikane et al., 1999; Laufer & Verzeano, 1967; Neuenschwander et al., 1999; Steinberg, 1966; Wachtmeister & Dowling, 1978; Yokoyama et al., 1964) and are potentially of clinical relevance (Adams & Dawson, 1971; Fortune, Bearse, Cioffi, & Johnson, 2002; Frishman et al., 2000; Rangaswamy, Hood, & Frishman, 2003; Wachtmeister, 1998). Experimental observation of coherent oscillations among retinal neurons may be confounded, however, by the lack of strong phase-locking to the stimulus onset and the possibility of wash-out in multi-trial averages due to random fluctuations in the underlying Fourier components (Stephens et al., 2006). Species differences, adaptation state and the level and type of anesthesia have also been reported to affect the amplitude of retinal oscillations (Doty & Kimura, 1963; Steinberg, 1966) and thus may confound experimental observations. Moreover, high frequency retinal oscillations are strongly size dependent (Ishikane et al., 1999; Neuenschwander et al., 1999; Stephens et al., 2006), precluding the use of standard visual stimuli that are matched to the preferred spatial frequency of the cells under study. *In vitro* preparations also lack long-range spatiotemporal correlations likely to result from high frequency ocular tremor (Martinez-Conde et al., 2004), which theoretical models indicate may tap into high-frequency resonant circuitry (Miller, Denning, George, Marshak, & Kenyon, 2006).

Separate from the issue of whether long-range spatiotemporal correlations can be evoked under physiological conditions and, if so, what additional information they might encode, there remains the question of how such information might be utilized by downstream elements. Both theoretical (Abeles, 1982; Kenyon, Fetz, & Puff, 1990; Kenyon, Puff, & Fetz, 1992) and experimental (Usrey, Alonso, & Reid, 2000) studies indicate that downstream neurons could operate as coincidence detectors, suggesting that spatiotemporal correlations of retinal origin directly influence the responses of targets elsewhere in the CNS (Castelo-Branco, Neuenschwander, & Singer, 1998; Doty & Kimura, 1963). Unsupervised, activity-dependent Hebbian plasticity rules, operating at individual synapses, have been used to train nodes in artificial neural networks to respond selectively to principal components (Hyvarinen, Karhunen, & Oja, 2001). Combined with non-linear dendritic sub-units (Gasparini & Magee, 2006; Oesch, Euler, & Taylor, 2005; Polsky, Mel, & Schiller, 2004), which may allow selective responses to pairs of synapses activated in close spatial and temporal proximity, there exists at least the possibility of extracting principal components defined by the spatiotemporal correlations present in short segments of spike train data. Sensitivity to oscillatory input, by exploiting resonances at the synaptic (Beierlein & Connors, 2002; Wespatat, Tennigkeit, & Singer, 2004), cellular (Vigh, Solessio, Morgans, & Lasater, 2003) and network levels (Miller et al., 2006; Rager & Singer, 1998), could extend the sensitivity to synchronous inputs so as to emphasize periodic temporal structure.

Finally, it is reasonable to speculate as to why the retina might have evolved a coding strategy based on extreme synergy in which local intensity information is distributed across many surrounding neurons. A possible clue is that to encode similar amounts of information using independent firing rates would require much higher levels of activity than are observed physiologically. We suggest that the primary advantage of a distributed code is to convey large amounts of information using a minimum number of spikes. Metabolic constraints have been previously suggested to influence the retinal code (Laughlin, de Ruyter van Steveninck, & Anderson, 1998). Given that the 2-point correlation function of natural scenes extends out to several degrees of visual angle (Reinagel & Zador, 1999), well beyond the central excitatory portion of a typical ganglion cell's central receptive field, it follows that individual retinal neurons are usually stimulated as part of larger visual features. The statistics of natural scenes therefore may permit the retina to distributed information among multiple neurons representing instantaneously defined ensembles of functionally related cells.

## Conclusions

The present findings suggest that the retina employs a highly distributed spatiotemporal code that utilizes extreme synergy, in which the firing events from any one neuron could only be properly interpreted in the context of multiple surrounding events. Such a code may have evolved to conserve energy by conveying maximal information using a minimal number of spikes.

## Acknowledgments

This research was supported by LDRD program at the Los Alamos National Laboratory. The author is deeply grateful to David Marshak, Ilya Nemenman and John Mosher for their comments and suggestions.

Commercial relationships: none.
Corresponding author: Garrett T. Kenyon.
Email: gkenyon@lanl.gov.
Address: Physics Division, P-21, P.O. Box 1663, MS D454, Los Alamos National Laboratory, Los Alamos, NM 87545.